\documentclass[english,aps,graphics,showpacs]{revtex4}
\usepackage{amssymb,amsmath,amsfonts,mathrsfs,eufrak,fancyhdr}
\usepackage{hyperref}
\usepackage{color}
\usepackage{makeidx,endnotes}
\usepackage{showidx}
\usepackage{theorem,eepic}
\usepackage{braket}

\def\d{\textrm{d}}

\def\<{\langle}
\def\>{\rangle}

\def\set#1{{\sf #1}}

\def\map#1{{\mathcal{#1}}}
\def\hil#1{{\mathscr{#1}}}

\def\Tr{\operatorname{Tr}}

\newtheorem{lemma}{Lemma}[section]

\newtheorem{corollary}[lemma]{Corollary}

\newtheorem{theorem}[lemma]{Theorem}

\def\Proof{\medskip\par\noindent{\bf Proof. }}
\newcounter{citac}[section]
\newcounter{index}

\begin{document}
\title{Asymptotics of quantum Markov processes: From algebraic structure to characterization of asymptotic states}
\author{J. Novotn\'y, J. Mary\v ska, and I. Jex}

\address{Department of Physics, Faculty of Nuclear Sciences and Physical Engineering,
Czech Technical University in Prague, B\v rehov\'a 7,
115 19 Praha 1 - Star\'e M\v{e}sto, Czech Republic
}

\date{\today}

\begin{abstract}
Markov processes play an important role in physics and the theory of open systems in particular. In this paper we study the asymptotic evolution of trace-nonincreasing homogenous quantum Markov processes (both types, discrete quantum Markov chains and continues quantum dynamical semigroups) equipped with a so-called strictly positive $\map T$-state in the Schr\"{o}dinger and the Heisenberg picture. We derive a fundamental theorem specifying the structure of the asymptotic and uncover a rich set of transformations between attractors of quantum Markov processes in both pictures. Moreover, we generalize the structure theorem derived for quantum Markov chains to quantum dynamical semigroups showing how the internal structure of generators of quantum Markov processes determines attractors in both pictures. Based on these results we provide two characterizations of all asymptotic and stationary states, both strongly reminding in form the well-known Gibbs states of statistical mechanics. We prove that the dynamics within the asymptotic space is of unitary type, i.e. quantum Markov processes preserve a certain scalar product of operators from the asymptotic space, but there is no corresponding unitary evolution on the original Hilbert space of pure states. Finally simple examples illustrating the derived theory are given.
\end{abstract}

\maketitle

\section{Introduction}
Any physical system is inevitably in contact with its surrounding.
Their mutual interaction may be unwanted or it is intended and
controlled but in all cases it breaks the unitary evolution of
individual subsystems \cite{Petrucione}. The resulting irreversible
open dynamics of the system of interest becomes highly involved and
in most cases escapes the possibility for analytical solutions. In
order to avoid the overall complexity of this issue some simplifying
assumptions are often applied. One of the most convenient approaches
focuses on quantum systems whose evolution can be described to some
extent by Markovian dynamics \cite{Zoller}. In such a case either
changes arising in the surrounding environment can be neglected or
are quickly dumped compared to the rate of mutual system-environment
interactions, or it is designed in the desired Markovian regime.
Basically two large classes of Markovian processes are at hand. Both
classes of quantum Markov processes (QMPs), continues quantum Markov
dynamical semigroups (QMDSs) \cite{Lindblad1976,Gorini1976} and
discrete quantum Markov chains (QMCHs) \cite{Gudder2008} are
frequently employed to investigate the equilibration of quantum
systems \cite{Alicki}, interaction of matter with electromagnetic
radiation \cite{Petrucione,Zoller} or decoherence effects in noisy
environment \cite{Joos2003}. On the other hand, engineered Markovian
dynamics provide a tool allowing for example the preparation of a
system in the desired quantum state, its storage and manipulation
\cite{Kraus,Kastoryano}, verification of quantum programs
\cite{Ying2013}, the synchronization of subsystems clocks \cite{Shi}
or environment assisted quantum transport
\cite{Rebentrostl,Stefanak}.

Despite the simplifications made towards the Markovian regime, a
full solution of quantum Markovian evolution constitutes in general
a challenging task. However, a large class of problems of
interest,like the already mentioned problem of decoherence protected
states, equilibrium states, transport efficiency or synchronization
of subsystems, may be resolved at the level of asymptotic dynamics.
Unlike to the rest of dynamics, the asymptotic part is fully
determined by the asymptotic (peripheral) spectrum of the relevant
Markovian generator and its associated eigenvectors (attractors)
forming the attractor (asymptotic) space. Understanding what
structures they form, their mutual algebraic properties or how they
are determined via internal structure of Markovian generators are of
central interest. Due to their relevance for quantum thermodynamics
and quantum cryptography, intensive research is devoted to the
analysis of fixed points of quantum Markov processes. Their
structure was analyzed for unital channels
\cite{Arias2002,Kribs2003} and for trace-preserving quantum dynamical
semigroups with a faithful invariant state \cite{Frigerio1982,Fagnola2004} and
lately for any quantum channel \cite{Kohout2010} and quantum
dynamical semigroups \cite{Gheondea2016}. The structure analysis of
the whole attractor space emerging from quantum Markov processes was
carried out for quantum operations equipped with a faithful
invariant state \cite{Novotny2012} and for trace-preserving quantum
dynamical semigroups \cite{Buamgartner2012,Albert2016}.

Another important issue we face concerns the mutual relationship
between attractors in the Schr\"{o}dinger and Heisenberg
picture. As the former contains all stationary states and the latter
all conserved quantities of the evolution, it plays a crucial role
in our understanding how conserved quantities determine the resulting
stationary state \cite{Albert2014}. In closed, unitary dynamics both
sets of attractors coincide and the same applies to quantum Markov
chains generated by unital channels \cite{Li2011}. However,
attractors in the Schr\"{o}dinger and the Heisenberg picture of
quantum Markov chains equipped with a faithful invariant state are
in general different and a simple algebraic relation between them
was presented in \cite{Novotny2012}.

In this paper we reveal that this relation \cite{Novotny2012} is
just a particular example from a whole family of mutual relations
among the two sets of attractors and we show that it applies to all
quantum Markov processes (including trace non-increasing processes
and quantum Markov dynamical semigroups) equipped with a so called
faithful $\map T$-state (a generalization of faithful invariant states for
trace non-increasing quantum Markov processes). This family is
generated by operator monotone functions and each its instance
provides a dual basis of the attractor space allowing to express the
asymptotic evolution for any initial state. We present two important
examples of operator monotone functions, each giving rise to a
useful (and convenient) characterization of stationary states (resp.
asymptotic states) via integrals of motion (resp. hermitian
attractors of quantum Markov process). Their importance is twofold.
First, it avoids problems with positivity of the dynamics induced
density operator which arises if we construct stationary or
asymptotic states directly from attractors. Second, these states
strongly resemble Gibbs states well known from statistical physics
\cite{Ballian}. We expect that this link might have further
applications in quantum processing.

The second goal of the paper is to examine how the inner structure
of Markovian generator determines attractor spaces of QMPs in the
Schr\"{o}dinger and the Heisenberg picture. Using relations between
attractors in both pictures we generalize previously known results
for QMCHs \cite{Novotny2012} and derive algebraic equations
determining attractors of continues QMDSs in both pictures in terms
of their Hamiltonian, Lindblad operators and eventually an optical
potential \cite{Alicki}. The obtained results apply to all discrete
QMCHs as well as continues QMDSs, which are equipped with a $\map
T$-state. We stress that these quantum Markov processes may not be
trace-preserving and therefore allow for analysis of physical
situations where part of the dynamics is not known \cite{Alicki} or
is gradually lost, e.g. evolution with a sink \cite{Stefanak}. Based
on the obtained algebraic equations we analyze the algebraic
structure of attractors in both pictures and specify the type of
evolution running inside the asymptotic space. In both, discrete and
continues, cases it is shown that the asymptotic dynamics is
reversible which with certain modifications might be seen as
unitary. In particular, we show that quantum channel capable to
reverse evolution inside the attractor space of QMCH is the
so-called Petz recovery map \cite{Jencova2012}. For continues QMDSs,
we derive a master equation driving the dynamics inside the
attractor space.

We briefly describe the structure of the paper. In section
\ref{sec:preliminaries} we provide settings and important
definitions that are used throughout the whole paper. The aim of
section \ref{sec:Markov_processes} is to introduce the concept of
quantum Markov processes, describe them in terms of their generators
and specify their asymptotic regime. Section \ref{sec:mutual
relations for attractors} is devoted to studies of the mutual
relationships between attractors of QMPs in the Schr\"{o}dinger and
the Heisenberg picture. Employing operator monotone functions we
construct dual basis for both, continues and discrete, QMPs. Finally
two important cases of operator monotone functions are discussed. In
section \ref{sec:Structure theorems for quantum Markov processes} we
prove the structure theorem for attractors of QMPs in both pictures
and reveal their algebraic properties. Two characterizations of
asymptotic and stationary states are given in section
\ref{sec:Asymptotic and stationary states of QMPs}. The description
of the dynamics inside the attractor space is investigated in
section \ref{sec:Dynamics within attractor spaces of QMPs}. Finally,
we examine two examples in section \ref{sec:examples} and conclude
in section \ref{sec:conclusion}.

\section{Preliminaries and definitions}
\label{sec:preliminaries}
Throughout the whole paper we assume a quantum system associated with a \textit{finite} $N$-dimensional Hilbert space $\hil H$ equipped with a scalar product $\<.,.\>$. Let $\set B(\hil H)$ be the associated Hilbert space of all operators acting on $\hil H$ and we denote its corresponding Hilbert-Schmidt product as $\left(A,B\right)=\Tr\{A^{\dagger}B\}$ and corresponding Hilbert-Schmidt norm $\parallel A \parallel_{HS} = \sqrt{\left(A,A\right)}$ with $A,B \in \set B(\hil H)$ ($A^{\dagger}$ denotes the adjoint operator of $A$).

A state of a quantum system is described by a density operator, a positive operator with unit trace. Let us denote the set of all states as $\map S(\hil H)$. The most general physical state change is given by a quantum operation $\map T: \set B(\hil H) \rightarrow \set B(\hil H)$, a linear completely positive map (CP) which admits decomposition into Kraus operators $\{A_j\}_{i=1}^{k} \subseteq  \set B(\hil H)$
\begin{equation}
\label{def:quantum_operation}
\map T\left(.\right) = \sum_{j=1}^{k} A_j \left(.\right) A_j^{\dagger}.
\end{equation}
Moreover, any quantum operation is supposed to be trace-nonincreasing and thus satisfying $T^{\dagger}(I) \leq I$ or equivalently expressed in terms of Kraus operators as $\sum\limits_{i=1}^{k} A_i^{\dagger} A_i \leq I$. Note that the adjoint map $\map T^{\dagger}$ of a quantum operation $T$ is also a completely positive map with Kraus operators $\{A_j^{\dagger}\}_{i=1}^{k}$, but it does not represent a quantum operation in general because the map $\map T^{\dagger}$ is not necessarily trace-nonincreasing. Let us introduce a little more terminology used in the context of quantum operations. If $\map T^{\dagger}(I)=\sum_j A_j^{\dagger} A_j = I$ we call the quantum operation $\map T$ a channel or a trace-preserving quantum operation. A quantum operation which leaves the maximally mixed state unchanged is called unital and satisfies $\map T(I)=\sum_j A_j A_j^{\dagger} = I$. A prominent example of unital channels are random unitary operations (random external fields). In the less restrictive cases when $\map P(I)=\sum_j A_j A_j^{\dagger} \leq I$ the quantum operation is called sub-unital.

From a different perspective quantum operations are linear maps acting on operators and for this reason they are called superoperators. Another examples of superoperators, we exploit in our analysis, are the left (resp. right) multiplication operator defined as $L_P(X) = PX$ (resp. $R_P(X) = XP$), where $P$ is a positive operator from $\set B(\hil H)$. From these two superoperators we can construct the relative modular operator $\Delta_{Q,P} = L_Q R_P^{-1}$, where $P$ is assumed to be strictly positive \cite{Ruskai1999}.

Assume a function $f$ defined on a real interval $\set I$ and a selfadjoint operator $A$ with its spectrum lying in the interval $I$. Invoking its spectral decomposition $A= U D U^{\dagger}$ with diagonal matrix $D =diag (\lambda_1, \lambda_2, \ldots )$ one can introduce the operator $f(A) = U f(D) U^{\dagger}$, where $f(D)$ is the diagonal matrix $f(D) = diag(f(\lambda_1), f(\lambda_2), \ldots)$. As will be shown a privileged role in the description of asymptotic behavior of QMPs is played by operator monotone functions. A function $k:[0,+\infty) \rightarrow [0,+\infty)$ is called an operator monotone if any two operators $A,B$ such that $A \geq B \geq 0$ implies $k(A) \geq k(B)$. Interestingly, L\"{o}wner has shown that all operator monotones can be characterized by an integral representation \cite{Lovner1934}.
\begin{theorem}
\label{theorem_integral_representation}
$k:[0,+\infty) \rightarrow [0,+\infty)$ is an operator monotone if and only if there is a positive finite measure $\mu$ such that
\begin{equation}
k(y) = \alpha + \beta y + \int_{0}^{+\infty} \frac{y(1+s)}{s+y} d\mu(s),
\end{equation}
with $\alpha = k(0)\geq 0$ and $\beta = \lim\limits_{t +\infty} \frac{f(t)}{t} \geq 0$.
\end{theorem}


\section{Homogeneous quantum Markov processes}
\label{sec:Markov_processes}
Let us recall the concept of markovian evolution and specify its form in quantum domain. In general, a stochastic process is called Markov if the future evolution of any present state is independent on its past. Its future is given solely by the action of a propagating map onto the present state. In the context of quantum mechanics it follows that the Markov evolution during any possible finite time interval $\<t_1,t_2\>$ is given by some quantum operation $\map T(t_1,t_2)$. Here we stress that we allow changes which are trace-nonincreasing. If any state change driven by quantum Markov process depends solely on the length of the time interval $\bigtriangleup t= t_2-t_1$, i.e.  $\map T(t_1,t_2)= \map T(\bigtriangleup t)$, the quantum Markov process is called homogenous. Thus to describe a state change under a homogenous quantum Markov process one does not even need to know the initial time of interval in which the state change takes place. In this work we investigate only homogenous quantum Markov processes and thus, to simplify notation, we use the notion quantum Markov process (QMP) to identify homogenous quantum Markov process.

We distinguish two classes of quantum Markov processes.  The first one are discrete quantum Markov chains (QMCHs). Their one step of evolution is governed by a generating quantum operation $\map T$ taking the state $\rho(n)$ emerging from previous $n$ iterations to the state $\rho(n+1) = \map T(\rho(n))$. Thus, within $n$ iterations the system initially prepared in the state $\rho(0)$ evolves into the state $\rho(n) = \map T^n(\rho(0))$. In the Heisenberg picture we have, instead of evolving states, evolving observables. A consistence of both descriptions requires that each initial observable $A(0)$ and initial state $\rho(0)$ fulfill mean-value condition at any step of QMCH
\begin{equation}
\label{def_consistence}
\< A(0)\>_{\map T^n(\rho(0))} = \< \map T^{'}_n (A(0))\>_{\rho(0)}.
\end{equation}
$T^{'}_n$ denotes the propagator describing the $n$ steps of QMCH evolution in the Heisenberg picture. From (\ref{def_consistence}) it follows that QMCH evolution in the Heisenberg picture is generated by the adjoint map $\map T^{\dagger}$, which is completely positive and unital.

The second class of continues QMPs is described by a quantum Markov dynamical (QMDS) semigroup of quantum operations $\map T_t$ transforming the state $\rho(t_1)$ at time $t_1$ into the state $\rho(t_2) = \map T_{t_2-t_1} \rho(t_1)$ at time $t_2$. Assuming uniformly continuous QMDSs, $\map T_t$ takes the form
\begin{equation}
\label{eq_QMDS}
\map T_t = \exp(\map L t).
\end{equation}
Due to Lindblad original work \cite{Lindblad1976} the generator $\map L: \set B(\hil H) \rightarrow \set B(\hil H)$ of QMDS can be written as
\begin{equation}
\label{eq_master}
\map L(X) = \map V(X) -KX- XK^{\dagger},
\end{equation}
with $\map V$ being a completely positive map with Kraus operators $\{L_j\}$ and K is an element of $\set B(\hil H)$. Splitting further $K$ into its hermitian and antihermitian part $K=iH + \frac{1}{2}\map V^{\dagger}(I) + G$ with Hamiltonian $H$ and an optical potential $G$ \cite{Alicki}, the generator of QMDS takes the form
\begin{equation}
\label{eq_QMDS_optical_potential}
\map L(X)=i[X,H] + \sum_j L_j X L_j^{\dagger} - \frac{1}{2} \left\{L_j^{\dagger}L_j,X \right\}-GX - XG.
\end{equation}
Here, the Lindblad operators $L_i$, Hamiltonian $H$ and the optical potential $G$ can be chosen arbitrarily, provided that $G$ is positive to ensure that the generated QMDS is trace-nonincreasing. The corresponding evolution under a QMDS is described by the Markov master equation
\begin{equation}
\frac{\d\rho(t)}{dt} = \map L(\rho).
\end{equation}
In case that QMDS $\map T_t$ is trace-preserving, we find $\map L^{\dagger}(I)= 0$ implying $G=0$ and the generator may be cast into the well known lindblad form for trace-preserving QMDSs
\begin{equation}
\label{lindbladian_form}
\map L(X)=i[X,H] + \sum_j L_j X L_j^{\dagger} - \frac{1}{2} \left\{L_j^{\dagger}L_j,X \right\}.
\end{equation}

Analogously as in the discrete case, the evolution of observables is given by the dynamical semigroup $\map T_t^{\dagger}$ whose generator takes the form
\begin{equation}
\label{eq_master_Heisenberg}
\map L^{\dagger}(A) = \map V^{\dagger}(A) -K^{\dagger} A- AK.
\end{equation}
If QMDS is trace-preserving the generator (\ref{eq_master_Heisenberg}) can be written with the help of the Hamiltonian as
\begin{equation}
\map L^{\dagger}(A)=-i[X,H] + \sum_j L_j^{\dagger} X L_j - \frac{1}{2} \left\{L_j^{\dagger}L_j,X \right\}.
\end{equation}

Quantum Markov processes are frequently used to model a simplified or effective evolution of open quantum systems. In comparison with the class of closed quantum evolutions an additional severe obstacle arises, if we start to analyze its evolution. This is due to the fact that both generators of QMPs do not commute with their adjoint map, i.e. they are neither hermitian nor normal. Consequently, this makes the generated dynamics much more involved and harder to solve as the standard method of spectral decomposition is not available. Let us list the implicit unpleasant consequences in details. First, such generator of a QMP may not be diagonalizable and then we are left with Jordan normal form given only in the basis of some generalized eigenvectors \cite{Horn1985}. Second, we miss a relationship between eigenvectors or generalized eigenvectors of the generator and its adjoint map. Consequently, corresponding (generalized) eigenvectors may be nonorthogonal and a construction of the important dual basis becomes a hard task.

However, if we are interested in the asymptotic dynamics of QMPs solely, the solution of its behavior is one step less complicated. Indeed, it has been shown \cite{Novotny2012} that the corresponding part of the generator responsible for asymptotic dynamics of QMP can always be diagonalized. The space $\set B(\hil H)$ can be decomposed into two parts, i.e. $\set B(\hil H)= \set{Atr} \oplus \set{Y}$ with attractor space $\set{Atr}$ supporting the asymptotic dynamics of a given QMP and a dying operator space $\set{Y}$ whose elements gradually vanish during the QMP. Both operator spaces $\set{Atr}$ and $\set{Y}$ are closed under the action of a given QMP. As the spectra of generators of discrete and continuous QMPs differ, definitions of their attractor spaces differ slightly as well. Let us start with discrete QMCHs in the Schr\"{o}dinger picture, where the attractor space is given as
\begin{equation}
\label{def_chain_attractors}
\set{Atr(\map T)}= \bigoplus_{\lambda \in \sigma_{as}} \set{Ker} \left(\map T - \lambda I \right),
\end{equation}
with asymptotic spectrum $\sigma_{as}$ containing all eigenvalues of the generator $\map T$ with modulo one. Assuming $X_{\lambda,i}$ form basis of each individual eigenspace $\set{Ker} \left(\map T - \lambda I \right)$ and $X^{\lambda,i}$ form the corresponding dual basis, i.e. $\left(X_{\lambda_1,i},X^{\lambda_2,j}\right)= \delta_{\lambda_1\lambda_2} \delta_{ij}$, we can write down the asymptotic dynamics of the given QMCH
\begin{equation}
\label{eq_chain_asymptotics}
\rho(n \gg 1) = \sum_{\lambda \in \sigma_{as},i} \lambda^n X_{\lambda,i} \Tr \left\{ \rho(0) \left(X^{\lambda,i}\right)^{\dagger}\right\}.
\end{equation}
In the Heisenberg picture the attractor space is given analogously as
\begin{equation}
\label{def_chain_attractors}
\set{Atr(\map T^{\dagger})}= \bigoplus_{\lambda \in \sigma_{as}} \set{Ker} \left(\map T^{\dagger} - \lambda I \right),
\end{equation}
and the asymptotic evolution of observables is written in the corresponding dual basis of the attractor space
\begin{equation}
\label{eq_chain_asymptotics_observ}
A(n \gg 1) = \sum_{\lambda \in \sigma_{as},i} \lambda^n \left(X^{\lambda,i}\right)^{\dagger} \Tr \left\{ A(0) X_{\lambda,i}\right\}.
\end{equation}
A special attention belongs to attractors associated with eigenvalue one. They are also called fixed points because they do not evolve during a given evolution. While fixed points of QMCH in the Schr\"{o}dinger picture $\set{Fix}(\map T)$ contain all stationary states, fixed points of QMCH in the Heisenberg picture $\set{Fix}(\map T^{\dagger})$ contain all integrals of motion. Both sets are nonempty and contain at least one positive operator.

Similarly, the attractor space of continuous QMPs is composed of the corresponding kernels of the generator $\map L$
\begin{equation}
\label{def_cont_attractors}
\set{Atr}= \bigoplus_{\lambda \in \sigma_{as}} \set{Ker} \left(\map L - \lambda I \right),
\end{equation}
where, in the continuous case, the asymptotic spectrum $\sigma_{as}$ contains only purely imaginary eigenvalues of the generator $\map L$. This can be deduced from the fact that $X$ is an eigenoperator of the generator $\map L$, i.e. $\map L(X)= aX$, if and only if $X$ is an eigenoperator of the quantum operation $\map T_t= \exp(\map L t)$ associated with the eigenvalue $\exp(at)$ for any positive time $t$. Hence, eigenoperator $X$ of $\map L$ associated with eigenvalue $a$ is an attractor of QMDS generated by $\map L$ iff $|\exp(at)|=1$ for any $t \geq 0$.
Provided we find the dual basis $X^{\lambda,i}$ for some chosen basis $X_{\lambda,i}$ of the attractor space, the asymptotic dynamics in the Schr\"{o}dinger picture takes the form
\begin{equation}
\label{eq_cont_asymptotics}
\rho(t \gg 1) = \sum_{\lambda \in \sigma_{as},i} \exp(\lambda t) X_{\lambda,i} \Tr \left\{ \rho(0) \left(X^{\lambda,i}\right)^{\dagger}\right\}.
\end{equation}

As we have already mentioned, compare to closed unitary evolution we miss a relationship between eigenvectors of the generator and its adjoint map. Therefore we can not be sure whether the operator spaces $\set{Atr}$ and $\set{Y}$ are mutually orthogonal and also different kernels forming the attractor space may be in general nonorthogonal. In addition finding the dual basis of the attractor space becomes a nontrivial problem. Thus, without such relation we loose a connection between the asymptotic dynamics of QMP in the Schr\"{o}dinger and the Heisenberg pictures.
One of our aims is to show that for a broad class of quantum Markov processes these obstacles can be removed and we can enjoy the benefits of an analogous theory we have got to use for closed unitary evolutions. Some relation between eigenvectors of the generator and its adjoint map were already revealed for quantum Markov chains \cite{Novotny2012}. Here we intend to show that there is a deep connection between eigenvectors of evolution generators in both pictures. The previously found results are an example of this connection and we generalize findings presented in \cite{Novotny2012} and extend their validity also to continues quantum Markov processes. As will be shown, it helps to uncover algebraic properties of attractors, especially algebraic properties of integrals of motion and stationary states of quantum Markov processes. Such results are essential for the complete understanding of equilibria and their formation.

\section{Relation between eigenvectors of QMPs in the Schr\"{o}dinger and the Heisenberg picture}
\label{sec:mutual relations for attractors}
This part is devoted to spectral properties of the generator responsible for the asymptotic dynamics of a given QMP. We first focus on quantum Markov chains, a generalization to continues quantum dynamical semigroups is straightforward.

\subsection{Quantum Markov chains}
\label{sec:QMCHs}
Throughout the rest of paper we assume a generating quantum operation $\map T$ equipped with a so-called faithful $\map T$-state $\sigma \in \set B(\hil H)$ which means that $\sigma$ is strictly positive and satisfies $\map T(\sigma) \leq \sigma$. This definition generalizes the concept of faithful invariant states \cite{Arias2002,Robinson} for QMPs which are not trace-preserving, i.e. they may decrease trace of input operators.
The central piece of the following considerations is the theorem \cite{Novotny2012} which establishes a basic relation between eigenvectors of quantum operation $\map T$ and its adjoint map.
\begin{theorem}
\label{theorem_original}
If $\lambda$ is from the asymptotic spectrum of quantum operation $\map T$ then for any attractor $X$ associated with the eigenvalue $\lambda$ we have
\begin{itemize}
\item[(1)] $X \in \set{Ker}\left(\map T - \lambda I  \right) \quad \Leftrightarrow \quad R_{\sigma^{-1}}(X) \in \set{Ker}\left(\map T^{\dagger} - \overline{\lambda} I\right)$,
\item[(2)] $X \in \set{Ker}\left(\map T - \lambda I  \right) \quad \Leftrightarrow \quad L_{\sigma^{-1}}(X) \in \set{Ker}\left(\map T^{\dagger} - \overline{\lambda} I\right)$.
\end{itemize}
\end{theorem}
Employing this theorem we can find a broad family of linear bijections mapping individual kernels $\set{Ker}\left(\map T - \lambda I  \right)$ of the attractor space onto itself. It can be formulated as follows.
\begin{theorem}
\label{theorem_bijections}
Let $\lambda$ be an element of asymptotic spectrum of quantum operation $\map T$ and $\sigma_1, \sigma_2$ two faithful not necessarily different $\map T$-states. Then any operator monotone function $k$ establishes a well defined bijection $k(\Delta_{\sigma_1,\sigma_2})$ onto attractors (in both the Schr\"{o}dinger and the Heisenberg picture) associated with the eigenvalue $\lambda$.
\end{theorem}
\Proof{ From theorem \ref{theorem_original} we can deduce that the relative modular operator $\Delta_{\sigma_1,\sigma_2}$ is strictly positive and defines a bijection onto the attractor subspace $\set{Ker}\left(\map T- \lambda I \right)$ as well as onto the attractor subspace $\set{Ker}\left(\map T^{\dagger}- \lambda I \right)$. This is true also for the map $sI + \Delta_{\sigma_1,\sigma_2}$ and its existing inversion, where $s$ is nonnegative and $I$ stands for the identity map. Moreover, all these maps commute mutually and thus for any finite measure $\mu(s)$ we observe that also the map
\begin{eqnarray}
k(\Delta_{\sigma_1,\sigma_2}) = \int_0^{+\infty} \frac{\Delta_{\sigma_1,\sigma_2}(1+s)}{s I +\Delta_{\sigma_1,\sigma_2}} d\mu(s) + \alpha I + \beta \Delta_{\sigma_1,\sigma_2}
\end{eqnarray}
with $\beta \geq 0$ is a strictly positive bijection onto the individual attractor subspaces $\set{Ker}\left(\map T- \lambda I \right)$ and $\set{Ker}\left(\map T^{\dagger}- \lambda I \right)$. According to theorem \ref{theorem_integral_representation} we conclude that any operator monotone function $k$ defines strictly positive bijection $k(\Delta_{\sigma_1,\sigma_2})$ onto these attractor subspaces in both pictures.
$~\Box$
}

We have found a broad and important family of linear bijections of attractor spaces. An apparent advantage of our construction is the fact that each operator monotone function defines a linear bijection of attractor spaces independently of the underlying Hilbert space, i.e. its dimension. In a similar way one can construct maps which are not bijections but map attractor spaces back into the same attractor space. Moreover, if we combine theorems \ref{theorem_original} and \ref{theorem_bijections} we receive a general relationship between attractors of QMCH in the Schr\"{o}dinger and Heisenberg picture.
\begin{corollary}
\label{corollary_eigenstate_relationship}
Let $\lambda$ be an element of the asymptotic spectrum of a quantum operation $\map T$ generating QMCH and $\sigma_1, \sigma_2$ its two faithful not necessarily different $\map T$-states. If $k$ is an operator monotone function then
\begin{equation}
X \in \set{Ker}\left(\map T - \lambda I\right) \quad \Leftrightarrow \quad \left(R_{\sigma_2}\right)^{-1}k(\Delta_{\sigma_1,\sigma_2}) (X) \in  \set{Ker}\left(\map T^{\dagger} - \overline{\lambda} I\right).
\end{equation}
\end{corollary}

Note that in corollary \ref{corollary_eigenstate_relationship} we could use also the superoperator $\left(R_{\sigma_1}\right)^{-1}k(\Delta_{\sigma_1,\sigma_2})$, which apparently does the same job. However, our choice is due to the advantage that maps $\left(R_{\sigma_2}\right)^{-1}$ and $k(\Delta_{\sigma_1,\sigma_2})$ commute and consequently their product is again a strictly positive operator. This allows us to define a new scalar product on the space $\set B(\hil H)$ for any choice of operator monotone function $k$
\begin{equation}
\left(X,Y\right)_k = \left(X,\left(R_{\sigma_2}\right)^{-1}k(\Delta_{\sigma_1,\sigma_2}) (Y)\right).
\end{equation}
Then we say that two operators $X$ and $Y$ are $k$-orthogonal if $\left(X,Y\right)_k =0$. Similarly, two sets in $\set B(\hil H)$ are $k$-orthogonal if any two elements from these two sets are mutually $k$-orthogonal. The direct consequence of the corollary  \ref{corollary_eigenstate_relationship} are the following $k$-orthogonality relations.
\begin{theorem}
\label{theorem_orthogonality}
Let $\lambda_1$ and $\lambda_2$ be different elements of the asymptotic spectrum of quantum operation $\map T$ generating QMCH and $\sigma_1, \sigma_2$ its two faithful not necessarily different $\map T$-states. If $k$ is an operator monotone function then
\begin{itemize}
\item[(1)] attractor subspaces $\set{Ker}\left(\map T - \lambda_1 I\right)$ and $\set{Ker}\left(\map T - \lambda_2 I\right)$ are $k$-orthogonal,
\item[(2)] attractor subspace $\set{Ker}\left(\map T - \lambda_1 I\right)$ and range $\set{Ran}\left(\map T - \lambda_1 I\right)$ are $k$-orthogonal.
\end{itemize}
\end{theorem}
\Proof{ Assume $X_i \in \set{Ker}\left(\map T - \lambda_i I\right)$. Then
\begin{eqnarray}
\left(X_1,X_2\right)_k &=& \frac{1}{\lambda_2}\left(X_1,\left(R_{\sigma_2}\right)^{-1}k(\Delta_{\sigma_1,\sigma_2}) (\map T(X_2))\right) \nonumber \\ &=& \frac{1}{\lambda_2}\left(\map T^{\dagger}\left( \left(R_{\sigma_2}\right)^{-1}k(\Delta_{\sigma_1,\sigma_2})(X_1)\right), X_2\right) \nonumber \\
&=& \frac{\lambda_1}{\lambda_2}\left( \left(R_{\sigma_2}\right)^{-1}k(\Delta_{\sigma_1,\sigma_2})(X_1), X_2\right) = \frac{\lambda_1}{\lambda_2}\left(X_1, X_2\right)_k \nonumber
\end{eqnarray}
following from the strict positivity of the operator $\left(R_{\sigma_2}\right)^{-1}k(\Delta_{\sigma_1,\sigma_2})$ and we find $\left(X_1,X_2\right)_k =0$.
In order to prove second statement consider $X \in \set{Ker}\left(\map T - \lambda_1 I\right)$ and $Y \in \set{Ran}\left(\map T - \lambda_1 I\right)$. Hence there exists $0 \neq Z \in  \set B(\hil H)$ such that $Y= \map T(Z) - \lambda Z$ and one can check
\begin{eqnarray}
\left(X,Y\right)_k &=& \left(X,\left(R_{\sigma_2}\right)^{-1}k(\Delta_{\sigma_1,\sigma_2}) \left(\map T(Z) - \lambda_1 Z \right)\right) \nonumber \\
&=& \left(\map T^{\dagger} \left(R_{\sigma_2}\right)^{-1}k(\Delta_{\sigma_1,\sigma_2})(X),Z\right) - \lambda \left(X,\left(R_{\sigma_2}\right)^{-1}k(\Delta_{\sigma_1,\sigma_2})(Z)\right) \nonumber \\
&=& \lambda \left[\left(\left(R_{\sigma_2}\right)^{-1}k(\Delta_{\sigma_1,\sigma_2})(X),Z\right) - \left(X,\left(R_{\sigma_2}\right)^{-1}k(\Delta_{\sigma_1,\sigma_2})(Z) \right) \right] = 0. \nonumber
\end{eqnarray}
$~\Box$
}

The theorem \ref{theorem_orthogonality} simply tells that attractors associated with different eigenvalues from asymptotic spectrum of generating quantum operation $\map T$ are $k$-orthogonal and they are also $k$-orthogonal to the in time dying space $\set{Y}$. In turn it means that we have found a dual basis $X^{\lambda,i}$ of eigenvectors $X_{\lambda,i}$. Assuming $k$ is a given operator monotone function, the dual basis $X^{\lambda,i}$ reads
\begin{equation}
\label{dual basis}
X^{\lambda,i} = \left(R_{\sigma_2}\right)^{-1}k(\Delta_{\sigma_1,\sigma_2})(X_{\lambda,i}) \left(X_{\lambda,i}, \left(R_{\sigma_2}\right)^{-1}k(\Delta_{\sigma_1,\sigma_2})(X_{\lambda,i})\right)^{-1}.
\end{equation}

\subsection{Quantum Markov Dynamical Semigroups}
\label{sec:QMDSs}
So far we have considered asymptotic evolution of discrete QMCHs. Let us show that the same theory applies to continues QMDSs as well. Apparently, it is sufficient to prove theorem \ref{theorem_original} for QMDSs. In order to proceed we need to define an analogy of faithful $\map T$-state for QMDSs.
We call a faithful state $\sigma$, i.e. $\sigma  > 0$, $\map T$-state if $\map T_t(\sigma)=\exp(\map T t)(\sigma) \leq \sigma$ holds for any positive time  $t$. This definition is more involved, especially when it comes to the point if a given QMDS satisfies it. However, in fact it is sufficient to check whether this condition is fulfilled for times from some right neighborhood of zero. Moreover, if the studied QMDS is trace-preserving, the definition of a faithful $\map T$-state $\sigma$ simply reduces to the condition $\map L(\sigma) = 0$. With this modification the theorem \ref{theorem_original} and all of the follow-up theory applies to QMDSs.
\begin{theorem}
\label{theorem_original_continues}
Let $\sigma$ be a faithful $\map T$-state of QMDS $\map T_t= \exp(\map L t)$.
If $\lambda$ is from the asymptotic spectrum of a QMDS $\map T_t$ then for any attractor $X$ associated with the eigenvalue $\lambda$ we have
\begin{itemize}
\item[(1)] $X \in \set{Ker}\left(\map L - \lambda I  \right) \quad \Leftrightarrow \quad R_{\sigma^{-1}}(X) \in \set{Ker}\left(\map L^{\dagger} - \overline{\lambda} I\right)$,
\item[(2)] $X \in \set{Ker}\left(\map L - \lambda I  \right) \quad \Leftrightarrow \quad L_{\sigma^{-1}}(X) \in \set{Ker}\left(\map L^{\dagger} - \overline{\lambda} I\right)$.
\end{itemize}
\end{theorem}
\Proof If $X \in \set{Ker}\left(\map L - \lambda I  \right)$ then $X \in \set{Ker}\left(\map T_t - \exp(\lambda t) I  \right)$ for any positive $t$. Due to theorem \ref{theorem_original} we have $R_{\sigma^{-1}}(X) \in \set{Ker}\left(\left(\map T_t\right)^{\dagger} - \overline{\exp(\lambda t)} I\right)$ for any positive $t$ which in turn, by differentiation with respect to time at $t=0$, means that $R_{\sigma^{-1}}(X) \in \set{Ker}\left(\map L^{\dagger} - \overline{\lambda} I\right)$. Other implications can be proven in the same way. $~\Box$

We have established a general theory for analyzing the attractor spaces of discrete and continues QMPs in both pictures. In the following we employ two examples of operator monotone functions which provide an additional insight into the inverse evolution restricted onto the asymptotic space and into the structure of asymptotic and stationary states.

\subsection{Operator monotone function $k(y)=y^{\alpha}$}
\label{sec:monotone_polynom}
 One of the well known operator monotones is $k(y) = y^{\alpha}$ for $\alpha \in (0,1]$. Its integral representation \cite{Bhatia1997} is given as
\begin{equation}
k(y) = y^{\alpha} = \int\limits_{0}^{+\infty} \frac{y s^{\alpha-1}}{y+ \alpha} \frac{\sin(\alpha \pi)}{\pi} ds.
\end{equation}

Taking into account theorem \ref{theorem_original} (or theorem \ref{theorem_original_continues} in case of QMDSs) the action of linear bijection onto individual subspaces composing the whole attractor space is given as $k(\Delta_{\sigma_1,\sigma_2})(X) = \sigma_1^{\alpha} X \sigma_2^{-\alpha}$ for any real $\alpha$. Consequently, for any real $\alpha$ the superoperator
\begin{equation}
\label{eq_bijection_one_half}
\left(R_{\sigma_2}\right)^{-1}k(\Delta_{\sigma_1,\sigma_2}) (X) = \sigma_1^{\alpha} X \sigma_2^{-\alpha -1}
\end{equation}
defines a one-to-one correspondence among attractors $\set{Ker}\left(\map T -\lambda I \right)$ and $\set{Ker}\left(\map T^{\dagger} -\overline{\lambda} I \right)$ (analogously for QMDSs). Attractors associated with different eigenvalues are mutually $k$-orthogonal with respect to the scalar product $\left(X,Y\right)_k = \left(X, \sigma_1^{\alpha} X \sigma_2^{-\alpha -1}(Y) \right)$.

Moreover, each of these new scalar products defines an associated adjoint map of the QO $\map T$. The case $\alpha = 1/2$ deserves our special attention. Indeed, choosing $\sigma_1$ and  $\sigma_2$ equal to  $\sigma$ we find that the  adjoint map to QO $\map T$ with respect to the scalar product $\left(X,Y\right)_{1/2} = \left(X,\sigma^{-1/2}Y\sigma^{-1/2}\right)$
takes the form
\begin{equation}
\label{eq_inverse_evolution}
\map T^{\ddagger}\left(.\right) = \sum_k \sigma^{1/2} A_k^{\dagger} \sigma^{-1/2} X \sigma^{-1/2}A_k\sigma^{1/2}.
\end{equation}
Apparently, this is again a completely positive trace-nonincreasing map. As will be shown later it is a quantum operation capable to reverse the evolution running inside the attractor space.

This type of bijections reveal an interesting structure of attractor spaces. One could naively infer that any faithful $\map T$-state must commute with all attractors, which would significantly decrease the  complexity of the attractor structure. However this is not true as we demonstrate in \ref{sec:examples}.

\subsection{Operator monotone function $k(y)=\log(1+y)$}
\label{sec:logarithm}
As another example we apply the derived theory to the operator monotone function $k(y)=\log(1+y)$ with its integral representation \cite{Bhatia1997}
\begin{equation}
k(y)=\log(1+y)= \int\limits_{1}^{+\infty} \frac{y}{y+s}ds.
\end{equation}
It follows that both operators $\log\left(I + \Delta_{\sigma_1,\sigma_2}\right)$ and $\log\left(I + \left(\Delta_{\sigma_1,\sigma_2}\right)^{-1}\right)$ are bijections which map individual attractor spaces, in both pictures, associated with a given eigenvalue onto itself. Employing the following identity
\begin{equation}
\log\left(I + \Delta_{\sigma_1,\sigma_2}\right)  = \log\left(I + \left(\Delta_{\sigma_1,\sigma_2}\right)^{-1}\right) + \log\left(\Delta_{\sigma_1,\sigma_2}\right)
\end{equation}
we find that the operator $\log\left(\Delta_{\sigma_1,\sigma_2}\right)$ is not necessary a bijection but also maps individual attractor spaces, in both pictures, back to the original individual attractor space. A straightforward calculation reveals that $\log\left(\Delta_{\sigma_1,\sigma_2}\right) = L_{\log(\sigma_1)} - R_{\log(\sigma_2)}$, which proves the following interesting statement.
\begin{corollary}
\label{theorem_logarithmic_endomorphism}
Let $\lambda$ be an element of the asymptotic spectrum of quantum operation $\map T$ and $\sigma_1, \sigma_2$ two faithful not necessarily different $\map T$-states. Then the map $L_{\log(\sigma_1)} - R_{\log(\sigma_2)}$ is an endomorphism onto the attractor space $\set{Ker}\left(\map T - \lambda I \right)$ of the QO $\map T$.
\end{corollary}

This statement provides a key ingredient for a characterization of all asymptotic states of QMPs (for details see section \ref{sec:Asymptotic and stationary states of QMPs}).

\section{Structure theorems for quantum Markov processes}
\label{sec:Structure theorems for quantum Markov processes}
This part is devoted to the analysis how an inner structure of a generator governs the attractors of the resulting QMPs. Thus in this part we presume that either Kraus operators $\{A_i\}$ of quantum operation generating QMCH or operators $\{L_i,H,G\}$ in \ref{eq_QMDS_optical_potential} defining QMDS are known. The ultimate goal is to uncover how these operators determine attractors of QMPs. The structure theorem for quantum Markov chains was already derived in \cite{Novotny2012}.
\begin{theorem}
\label{theorem:attractors of QMCH}
Let $\map T: \set B(\hil H) \rightarrow \set B(\hil H)$ be a quantum operation (\ref{def:quantum_operation}) equipped with a faithful $\map T$-state $\sigma$. If $X$ is an attractor of QMCH in the Schr\"{o}dinger picture generated by $\map T$ associated with eigenvalue $\lambda$ then it necessary satisfies the following set of equations
\begin{eqnarray}
A_j X \sigma^{-1} &=& \lambda X \sigma^{-1} A_j, \qquad \qquad A_j^{\dagger} X \sigma^{-1} = \overline{\lambda} X \sigma^{-1} A_j^{\dagger} \nonumber \\
A_j \sigma^{-1}X  &=& \lambda \sigma^{-1}X A_j, \qquad \qquad A_j^{\dagger} \sigma^{-1}X  = \overline{\lambda} \sigma^{-1}X A_j^{\dagger}
\label{eq_attractors_discrete}
\end{eqnarray}
for all $j$'s. If $X$ is an attractor of QMCH in the Heisenberg picture associated with eigenvalue $\overline{\lambda}$ then it necessary satisfies the set of equations (\ref{eq_attractors_discrete}) for $\sigma = I$. \\
Moreover, if quantum operation $\map T$ is either trace-preserving or $\map T$-state $\sigma$ is additionally stationary then the reverse statement applies as well.
\end{theorem}
The importance of the theorem \ref{theorem:attractors of QMCH} is twofold. First, it significantly simplifies the calculation of asymptotic behavior of QMCHs. We should stress that this can be done analytically in many cases, especially if the studied evolution poses some sort of symmetry. Second, it also reveals the algebraic structure of attractors and we shall discuss this point simultaneously for discrete and continuous quantum Markov evolution later.
In the following we show that QMDSs follow a similar structure theorem for their attractors.
\begin{theorem}
\label{theorem:attractors of QMDS}
Let $\map T_t: \set B(\hil H) \rightarrow \set B(\hil H)$ be a quantum Markov dynamical semigroup with generator $\map L$ (\ref{eq_QMDS_optical_potential}) equipped with a faithful $\map T$-state $\sigma$. If $X \in \set B(\hil H)$ is an attractor of QMDS in the Schr\"{o}dinger picture associated with eigenvalue $\lambda = ia$ then the following set of equations holds
\begin{eqnarray}
\label{commutation_lindblad}
  [L_j,X\sigma^{-1}] &=& [L_j,\sigma^{-1}X]=[L_j^{\dagger},X\sigma^{-1}]=[L_j^{\dagger},\sigma^{-1}X]= 0, \\
 \label{commutation_optical_potential}
 [X\sigma^{-1},G] &=& [\sigma^{-1}X,G] = 0, \\
\label{commutation_Hamiltonian}
 [\sigma^{-1}X,H] &=& a\sigma^{-1}X, \quad [X\sigma^{-1},H] = aX\sigma^{-1}
\label{eq_K_solution}
\end{eqnarray}
for all $j$'s. If $X \in \set B(\hil H)$ is an attractor of QMDS in the Heisenberg picture associated with eigenvalue $\lambda = -ia$ then it must satisfy all equations (\ref{commutation_lindblad}), (\ref{commutation_optical_potential}) and (\ref{commutation_Hamiltonian}) with $\sigma = I$. \\
If QMDS $\map T_t$ is either trace-preserving or $\map T$-state $\sigma$ is stationary then the reverse statement applies as well.
\end{theorem}
\Proof
We first derive the necessary conditions which follow from the assumption that operator $X$ is an attractor of QMDS, i.e. $\map L(X)= iaX$. The generator $\map L$ maps hermitian operators back to hermitian operators and thus also $X^{\dagger}$ is an attractor, i.e. $\map L(X^{\dagger})= iaX^{\dagger}$. Employing theorem \ref{theorem_original_continues} we find that
\begin{equation}
\label{additional attractors}
\map L^{\dagger}\left(\sigma^{-1}X\right)= -ia\sigma^{-1}X, \quad \map L^{\dagger}\left(\sigma^{-1}X^{\dagger}\right)= ia\sigma^{-1}X^{\dagger}, \quad L^{\dagger}\left(X^{\dagger}\sigma^{-1}\right)= iaX^{\dagger}\sigma^{-1},
\end{equation}
which can be equivalently rewritten as
\begin{eqnarray}
\sum_i L_i^{\dagger} \sigma^{-1}X L_i &=& -ia\sigma^{-1}X + K^{\dagger}\sigma^{-1}X + \sigma^{-1}XK, \nonumber \\
\sum_i L_i^{\dagger} \sigma^{-1}X^{\dagger} L_i &=& ia\sigma^{-1}X^{\dagger} + K^{\dagger}\sigma^{-1}X^{\dagger} + \sigma^{-1}X^{\dagger}K, \nonumber \\
\sum_i L_i^{\dagger} \sigma^{-1}X^{\dagger} L_i &=& ia\sigma^{-1}X^{\dagger} + K^{\dagger}\sigma^{-1}X^{\dagger} + \sigma^{-1}X^{\dagger}K, \nonumber \\
\sum_i L_i^{\dagger}L_i &\leq& K + K^{\dagger}.
\label{ineqalities_1}
\end{eqnarray}
The last inequality expresses the fact that QMDS is trace-nonincreasing. In order to proceed we still need one more inequality. From theorem \ref{theorem_original_continues} we have for each positive $t$
\begin{equation}
\label{additional attractors}
\map T_t^{\dagger}\left(\sigma^{-1}X\right)= e^{-iat}\sigma^{-1}X, \quad T_t^{\dagger}\left(X^{\dagger}\sigma^{-1}\right)= e^{iat}X^{\dagger}\sigma^{-1}. \nonumber
\end{equation}
Using Schwartz operator inequality \cite{Bhatia,Paulsen} for subunital QOs $\map T_t^{\dagger}$ we obtain
\begin{equation}
\map T_t^{\dagger} \left(\sigma^{-1}XX^{\dagger}\sigma^{-1} \right) \leq \map T_t^{\dagger}\left(\sigma^{-1}X\right) T_t^{\dagger}\left(X^{\dagger}\sigma^{-1}\right) = \sigma^{-1}XX^{\dagger}\sigma^{-1}, \nonumber
\end{equation}
As this applies to all positive $t$ we find $\map L^{\dagger}\left(\sigma^{-1}XX^{\dagger}\sigma^{-1}\right) \leq 0$ or equivalently
\begin{equation}
\sum_i L_i^{\dagger} \sigma^{-1}XX^{\dagger}\sigma^{-1} L_i \leq K^{\dagger}\sigma^{-1}XX^{\dagger}\sigma^{-1} + \sigma^{-1}XX^{\dagger}\sigma^{-1} K.
\label{inequalities_2}
\end{equation}
Let us set $V_i = X\sigma^{-1}L_i - L_i X\sigma^{-1}$. Equipped with relations (\ref{ineqalities_1}) and (\ref{inequalities_2}) we receive
\begin{eqnarray}
\sum_i V_i^{\dagger}V_i &=& \sum_i L_i^{\dagger} \sigma^{-1}XX^{\dagger}\sigma^{-1} L_i - \left(\sum_i L_i^{\dagger}\sigma^{-1}XL_i\right)X\sigma^{-1} \nonumber \\
&-& \sigma^{-1}X^{\dagger}\left(\sum_i L_i^{\dagger}X\sigma^{-1}L_i\right) + \sigma^{-1}X^{\dagger}\left(\sum_i L_i^{\dagger}L_i \right) X \sigma^{-1} \leq 0. \nonumber
\end{eqnarray}
This inevitably means that all operators $V_i$ are equal to zero and consequently $[L_i,X\sigma^{-1}]=0$. Due to theorem \ref{theorem_original_continues} the operator $\tilde{X}=\sigma^{-1}X\sigma$ is an attractor satisfying $\map L\left(\tilde{X}\right)=ia\tilde{X}$ which proves the commutation relation $[L_i,\tilde{X}\sigma^{-1}]=[L_i,\sigma^{-1}X]=0$. Both sets of these commutation relations are also valid for the attractor $X^{\dagger}$. Taking adjoint of these equations we obtain the last two commutation relations $[L_i^{\dagger},X\sigma^{-1}]=[L_i^{\dagger},\sigma^{-1}X]=0$.

In order to prove commutation relations (\ref{commutation_optical_potential}) and (\ref{commutation_Hamiltonian}) we will first derive commutation relations with the operator $K$ for which we have to prove two useful equalities. Using commutation relations (\ref{commutation_lindblad}) we can rewrite equation $\map L(X) = iaX$ into the form
\begin{equation}
Z_1 \equiv X\sigma^{-1}KX^{\dagger} - KX\sigma^{-1}X^{\dagger} - iaX\sigma^{-1}X^{\dagger}= X\sigma^{-1}\left[\sum_i K\sigma + \sigma K^{\dagger} - L_i\sigma L_i^{\dagger}  \right]\sigma^{-1}X^{\dagger}. \nonumber
\end{equation}
As $\sigma$ is a faithful $\map T$-state, it follows that operator $Z_1$ is positive. On the other hand, using
\begin{equation}
\map L^{\dagger}\left(X\sigma^{-1}\right) = \sum_i L_i^{\dagger}X \sigma^{-1}L_i - K^{\dagger}X\sigma^{-1} - X\sigma^{-1}K = -iaX\sigma^{-1}, \nonumber
\end{equation}
we find
\begin{eqnarray}
\Tr Z_1 &=& \Tr\left\{\left[\sum_i L_i^{\dagger}L_i - K^{\dagger} - K\right]X\sigma^{-1}X^{\dagger}\right\} = \Tr\left\{\map L^{\dagger}(I)X\sigma^{-1}X^{\dagger} \right\} \nonumber \\ &=& \Tr\left\{\map L\left( X\sigma^{-1}X^{\dagger} \right) \right\} \leq 0. \nonumber
\end{eqnarray}
A positive operator with a non-positive trace must be equal to the zero operator, i.e. $Z_1=0$.

In order to obtain the second required equality we start from the equation $\map L^{\dagger}\left(X\sigma^{-1} \right) = -iaX\sigma^{-1}$. Employing (\ref{commutation_lindblad}) we find that
\begin{equation}
 Z_2 \equiv \sigma^{-1}X^{\dagger}KX\sigma^{-1} - \sigma^{-1}X^{\dagger}X\sigma^{-1}K + ia\sigma^{-1}X^{\dagger}X\sigma^{-1} = \sigma^{-1}X^{\dagger}\left[\sum_i K + K^{\dagger} - L_iL_i^{\dagger}  \right]X\sigma^{-1}. \nonumber
\end{equation}
Operator $Z_2$ is obviously positive but on the other hand its trace can be rewritten using
\begin{equation}
\map L^{\dagger}\left(\sigma^{-1}X^{\dagger}\right) = \sum_i L_i^{\dagger}\sigma^{-1}X^{\dagger} L_i - K^{\dagger}\sigma^{-1}X^{\dagger} - \sigma^{-1}X^{\dagger}K = ia\sigma^{-1}X^{\dagger}, \nonumber
\end{equation}
into the inequality
\begin{eqnarray}
\Tr Z_2 &=& \Tr\left\{\left[\sum_i L_i^{\dagger}L_i - K^{\dagger} - K\right]\sigma^{-1}X^{\dagger}X\sigma^{-1}\right\} = \Tr\left\{\map L^{\dagger}(I)\sigma^{-1}X^{\dagger}X\sigma^{-1} \right\} \nonumber \\ &=& \Tr\left\{\map L\left( \sigma^{-1}X^{\dagger}X\sigma^{-1} \right) \right\} \leq 0. \nonumber
\end{eqnarray}
Hence we conclude that $Z_2=0$.

Assume now operator $W=KX\sigma^{-1/2}-X\sigma^{-1}K\sigma^{1/2}+iaX\sigma^{-1/2}$. Based on the obtained equalities its Hilbert-Schmidt norm can be expressed as
\begin{eqnarray}
\parallel W \parallel_{HS}^2 &=& \Tr\left\{X\sigma^{-1}K\sigma K^{\dagger}\sigma^{-1}X^{\dagger} - KXK^{\dagger}\sigma^{-1}X^{\dagger} - iaXK^{\dagger}\sigma^{-1}X^{\dagger}  \right\} \nonumber \\
&+& \Tr\left\{KX\sigma^{-1}KX^{\dagger}K^{\dagger} - X\sigma^{-1}KX^{\dagger}K^{\dagger} + iaX\sigma^{-1}X^{\dagger}K^{\dagger}  \right\} \nonumber \\
&+& \Tr\left\{iaX\sigma^{-1}KX^{\dagger} - iaKX\sigma^{-1}X^{\dagger}K^{\dagger} + a^2X\sigma^{-1}X^{\dagger}\right\} \nonumber \\
&=& -\Tr\left\{Z_2\sigma K^{\dagger} \right\} - \Tr\left\{Z_1K^{\dagger} \right\} +ia\Tr\left\{Z_1\right\} = 0. \nonumber
\end{eqnarray}
Thus we get $W=0$ yielding the commutation relation $\left[X\sigma^{-1},K  \right]=iaX\sigma^{-1}$. This commutation relation must be valid also for attractors $\sigma^{-1}X\sigma$, and $X^{\dagger}$ which provides commutation relations $\left[\sigma^{-1}X,K  \right]=ia\sigma^{-1}X$, $\left[X\sigma^{-1},K^{\dagger}\right]= -iaX\sigma^{-1}$ and $\left[\sigma^{-1}X,K^{\dagger}  \right]=-ia\sigma^{-1}X$. Now using $K=iH + \frac{1}{2}\map V^{\dagger}(I) + G$ we finally arrive at commutation relations (\ref{commutation_optical_potential}) and (\ref{commutation_Hamiltonian}).

If QMDS is trace-preserving the commutation relations (\ref{commutation_lindblad}) and (\ref{commutation_Hamiltonian}) constitute sufficient conditions for $X$ being an attractor. Indeed, a straightforward calculation shows
\begin{eqnarray}
\map L^{\dagger}\left(X\sigma^{-1}\right) &=& \sum_i L_i^{\dagger} X\sigma^{-1} L_i - K^{\dagger}X\sigma^{-1} - X\sigma^{-1}K \nonumber \\
&=& X\sigma^{-1} \left[\sum_i L_i^{\dagger} L_i - K^{\dagger} - K \right] - iaX\sigma^{-1}\nonumber \\
&=& X\sigma^{-1}\map L^{\dagger}(I) - iaX\sigma^{-1} = -iaX\sigma^{-1}. \nonumber
\end{eqnarray}
Hence $X\sigma^{-1}$ is an attractor in the Heisenberg picture and consequently due to theorem \ref{theorem_original_continues} is $X$ an attractor satisfying $\map L(X)= iaX$.

Similarly, if faithful $\map T$-state $\sigma$ is stationary, i.e. $\map L(\sigma)=0$, we get due to (\ref{commutation_lindblad}) and (\ref{eq_K_solution})
\begin{eqnarray}
\map L(X) = \sum_i L_iXL_i^{\dagger} - KX - XK^{\dagger} = X\sigma^{-1}\map L(\sigma) +iaX = iaX, \nonumber
\end{eqnarray}
confirming that $X$ is an attractor following $\map L(X) = iaX$.

Analogous statements for attractors in the Heisenberg picture follow directly from  \ref{theorem_original_continues}. Assuming $\map L^{\dagger}(X) = iaX$ we have $\map L\left(X\sigma\right)=-iaX\sigma$ and $\map L\left(\sigma X\right)=-ia\sigma X$ and thus equations (\ref{commutation_lindblad}), \ref{commutation_optical_potential} and (\ref{eq_K_solution}) have to be fulfilled for operators $X\sigma$ and $\sigma X$.
$~\Box$

Note that Lindblad operators and the optical potential are involved in the selection of attractors only. The corresponding asymptotic spectrum is fully determined by Hamiltonian.

Both structure theorems provide an insight into the algebraic structure of attractor spaces. It is already well known that if $X$ is an eigenvector of QMP associated with eigenvalue $\lambda$ then $X^{\dagger}$ is an eigenvector associated with eigenvalue $\overline{\lambda}$ \cite{Bhatia}. Employing theorems \ref{theorem:attractors of QMCH} and \ref{theorem:attractors of QMDS} it follows for trace-preserving QMP or QMP with stationary faithful $\map T$-state $\sigma$ that if  $X_1$ and $X_2$ are attractors of QMP associated with eigenvalue $\lambda_1$ and $\lambda_2$ then $X_1 \sigma^{-1}X_2$, or any permutation of these three operators, is also attractor associated with eigenvalue $\lambda_1 \lambda_2$. Similarly, we have the same statement for attractors in the Heisenberg picture with $\sigma = I$. Consequently, while attractors in the Schr\"{o}dinger picture do not form algebra, for attractors in the Heisenberg picture we can formulate the following corollary. \begin{corollary}
\label{corollary_algebra}
Assume a QMP which is either trace-preserving or whose faithful $\map T$-state $\sigma$ is stationary. Then the whole attractor space in the Heisenberg picture and its subspace of fixed points form $C^{*}$ algebra.
\end{corollary}
This statement combined with theorem \ref{theorem_bijections} gives us a useful characterization of all stationary states or even all asymptotic states.

\section{Asymptotic and stationary states of QMPs}
\label{sec:Asymptotic and stationary states of QMPs}
The description of the asymptotics in terms of attractors is elegant. However we have to face the fact that the attractors are not states.
They constitute building blocks, operators, from which asymptotic states (\ref{eq_chain_asymptotics}) and (\ref{eq_cont_asymptotics}) are constructed. The range of coefficients $\Tr \left\{ \rho(0) \left(X^{\lambda,i}\right)^{\dagger}\right\}$ involved in these formulas is largely unknown and makes a complete characterization of asymptotic states quite involved and in many cases unfeasible. In this part we present two characterizations of asymptotic states and its subset of stationary states of quantum trace-preserving Markov processes allowing a glance on the asymptotics.

The first characterization relies on $C^{*}$ algebraic structure of attractors in the Heisenberg picture. Obviously, the exponential map
\begin{equation}
\exp(A) = \sum_{k=0}^{+\infty}\frac{A^k}{k!}
\end{equation}
maps any hermitian operator $A$ from the attractor space in the Heisenberg picture (resp. any integral of motion $A$) to a strictly positive operator from the attractor space in the Heisenberg picture (resp. to a strictly positive integral of motion). Using the bijection (\ref{eq_bijection_one_half}) with $\alpha= -1/2$ between attractors in both pictures we find for any hermitian operator $A$ from the attractor space in the Heisenberg picture (resp. any integral of motion $A$) that the operator
\begin{equation}
\label{exponential_form}
\sigma^{1/2} \exp(A) \sigma^{1/2}
\end{equation}
corresponds, to a strictly positive asymptotic state (resp. to a strictly positive stationary state). Because the identity operator is an integral of motion we can always choose operator $A$ in a way that (\ref{exponential_form}) is properly normalized.

However we have a more ambitious inverse task in mind, namely  to show that any asymptotic state (resp. any stationary state) may be written as (\ref{exponential_form}). To proceed we employ the following analytic formula for the logarithmic map
\begin{equation}
\log(I+A) = \sum_{k=1}^{+\infty} \frac{(-1)^{k+1}}{k}A^k,
\end{equation}
which converges for any hermitian operator with its spectrum within the interval $\left(-1,1\right]$. Assume a strictly positive asymptotic state (resp. a strictly positive stationary state) $\rho$. Then $\omega=\gamma\sigma^{-1/2} \rho \sigma^{-1/2}$ is a strictly positive operator from the attractor space in the Heisenberg picture (resp. a strictly positive integral of motion) normalized by $\gamma= 1/\Tr\left(\sigma^{-1/2} \rho \sigma^{-1/2}\right)$. Consequently,
\begin{equation}
\label{logarithm_of_asymptotic}
\log(\omega) = \log(I + (\omega- I)) = \sum_{k=1}^{+\infty} \frac{(-1)^{k+1}}{k}(\omega-I)^k
\end{equation}
is well defined and yields a hermitian operator from the attractor space in the Heisenberg picture (resp. an integral of motion). Indeed, the identity $I$ is an integral of motion ensuring trace-preservation of the given QMP and the attractor space as well as the space of fixed points of quantum Markov evolution in the Heisenberg picture is closed under all algebraic operations involved in (\ref{logarithm_of_asymptotic}). Thus any strictly positive asymptotic state takes the form (\ref{exponential_form}). We can generalize this statement also to all asymptotic states, because strictly positive asymptotic states constitute a dense set inside all asymptotic states. Indeed, starting with an asymptotic state $\rho$, we can define a set of strictly positive asymptotic states
\begin{equation}
\omega(s) \equiv (1-s) \rho + s\sigma = \sigma^{1/2} \exp\left(A(s)\right) \sigma^{1/2}
\end{equation}
with $s \in (0,1]$.
Apparently,
\begin{equation}
\rho = \lim_{s \rightarrow 0_{+}} \omega(s) = \lim_{s \rightarrow 0_{+}} \sigma^{1/2}\exp\left(A(s)\right) \sigma^{1/2}.
\end{equation}
A convenient way to express this limit is to choose a hermitian base $\left\{Z_i\right\}$ (with $Z_i^{\dagger}=Z_i$) of all asymptotic operators. As the attractor space in the Heisenberg picture is enclosed under the adjoint map, such basis always exist. Hence any asymptotic state $\rho$ can be written as
\begin{equation}
\label{limit_exponential_form}
\rho = \lim_{s \rightarrow 0_{+}} \sigma^{1/2} \exp\left(\sum_i \beta_i(s) Z_i\right)\sigma^{1/2}.
\end{equation}
Note that if $\{Z_i\}$ constitute a hermitian basis of fixed points of quantum Markov evolution in the Heisenberg picture then (\ref{limit_exponential_form}) provides us with all stationary states of the given QMP.
In fact, the limiting procedure (\ref{limit_exponential_form}) means that some of these real coefficients $\beta_i(s)$ approach, in the limit $\lim_{s \rightarrow 0_{+}}$, either plus or minus infinity, otherwise the state $\rho$ is strictly positive. This might appear counterintuitive, but in statistical physics we meet this situation frequently. For example, one obtains a ground state of a canonical ensemble by taking the temperature limit $T \rightarrow 0_{+}$ which corresponds to $\beta \equiv 1/(kT) \rightarrow +\infty$ \cite{Ballian}.

The second characterization of asymptotic and consequently also stationary states provides corollary  \ref{theorem_logarithmic_endomorphism}. Assume that $\sigma$ is a faithful invariant state and $\rho$ is some strictly positive stationary state, i.e. $\rho$ is an attractor associated with eigenvalue one but it is also a $\map T$-state. According to the corollary the operator $\rho\log(\rho) - \rho \log(\sigma)$ is an attractor associated with eigenvalue one. Hence, the operator $\log(\rho) - \log(\sigma)$ is an hermitian attractor of evolution in the Heisenberg picture associated with eigenvalue one, i.e. it is an integral of motion. By choosing a hermitian base $\{ Y_i\}$ of integrals of motion we can write down any strictly positive stationary state $\rho$ into the form
\begin{equation}
\label{eq_exp_stationary states}
\rho = \exp\left(\log(\sigma) + \sum_i \gamma_i Y_i  \right),
\end{equation}
where $\gamma_i$ are real expansion coefficients of the operator $\log(\rho) - \log(\sigma)$ in the base $\{ Y_i\}$.

A generalization of (\ref{eq_exp_stationary states}) to all strictly positive asymptotic states follows from the fact that all asymptotic states (\ref{eq_chain_asymptotics}) (resp. (\ref{eq_cont_asymptotics}) in continues case) are actually stationary states of the quantum operation
\begin{equation}
\tilde{\map T}(\rho) = \sum_{\lambda \in \sigma_{as},i} X_{\lambda,i} \Tr \left\{ \rho \left(X^{\lambda,i}\right)^{\dagger}\right\}.
\end{equation}
Indeed, as the asymptotic spectrum $\sigma_{as}$ of QMP contains a finite number of eigenvalues, we can choose an ascending sequence of natural numbers $n_j$ in such a way \cite{Wolf} that
\begin{equation}
\tilde{\map T} = \lim_{j \rightarrow +\infty} \map T^{n_j} \quad {\rm resp. } \quad \tilde{\map T} = \lim_{j \rightarrow +\infty} \map \exp\left(\map L n_j\right).
\end{equation}
Therefore $\tilde{\map T}$ is quantum operation which projects all states onto the set of asymptotic states of the original QMP. All attractors of the original QMP are fixed points of $\tilde{\map T}$ and because of positivity of the original QMP we have also $\tilde{\map T}(\sigma) \leq \sigma$. Now, let $\rho$ be a strictly positive asymptotic state of the original QMP. It is stationary state of $\tilde{\map T}$ and thus it can be written as (\ref{eq_exp_stationary states}),
where $\{Y_i\}$ is chosen as a hermitian base of fixed points of $\tilde{\map T}$ and consequently it forms a hermitian base of the attractor space of the original QMP. Thus
\begin{equation}
\label{eq_exp_asymptotic states}
\rho = \exp\left(\log(\sigma) + \sum_i \gamma_i Z_i  \right),
\end{equation}
with a hermitian basis $\{Z_i\}$ of the attractor space in the Heisenberg picture, describes all strictly positive asymptotic states of the given QMP. Following the same recipe as in the first characterization we finally enlarge (\ref{eq_exp_stationary states}) to all asymptotic states of the given trace-preserving QMP.

We also stress that while the first characterization (\ref{limit_exponential_form}) is valid only for trace-preserving QMPs, the second characterization of asymptotic states (\ref{eq_exp_asymptotic states}) applies to trace-nonincreasing QMPs as well provided there is a stationary strictly positive state $\sigma$. We have found two expressions for asymptotic states of QMPs. They are, in general, different as we show in examples \ref{sec:examples}. We expect that especially the asymptotic form (\ref{eq_exp_stationary states}) can be further exploited to study thermodynamic properties of QMPs. It has also the advantage that it may apply to trace-nonincreasing QMPs also. Both forms of asymptotic states (\ref{limit_exponential_form}) (\ref{eq_exp_stationary states}) remind of Gibbs states, the well known family of macroscopic states in statistical physics \cite{Ballian}. A detailed investigation of their intricate connection will be presented elsewhere.

\section{Dynamics within attractor spaces of QMPs}
\label{sec:Dynamics within attractor spaces of QMPs}
From a different perspective the attractor space is the part of the total Hilbert space $\set B(\hil H)$ which is exempt from effects of decay, decoherence and dissipation in dependence on the detailed features of the process. In principle, any information encoded in this subspace should be fully retrieved. In order to recover this information we need to understand in detail the type of evolution rules the asymptotic states of QMPs. It was shown \cite{Novotny2009} that asymptotic states of random unitary operations undergo an unitary evolution. In general case of QMPs it is stated \cite{Baumgartner2012,Kohout2010,Albert2016} that the evolution inside the attractor space should be of unitary type as well, because all attractors during such evolution acquire only its individual phase (\ref{eq_chain_asymptotics},\ref{eq_cont_asymptotics}). Using structure theorems (\ref{theorem:attractors of QMCH}) and (\ref{theorem:attractors of QMDS}) we show that for trace-preserving QMPs equipped with a faithful invariant state it is in some sense true.

We start with discrete QMCHs. Let $X$ be any operator from the attractor space in the Schr\"{o}dinger picture, i.e. $X = \sum\limits_{\lambda,i} c_{\lambda,i} X_{\lambda,i}$. A straightforward calculation exploiting a quantum operation (\ref{eq_inverse_evolution}) gives us
\begin{eqnarray}
\map T^{\ddagger} \map T(X) &=& \map T^{\ddagger}\left(\sum_{\lambda,i}c_{\lambda,i}\sum_kA_kX_{\lambda,i}\sigma^{-1}\sigma A_k^{\dagger} \right) = \map T^{\ddagger}\left(\sum_{\lambda,i}\lambda c_{\lambda,i}X_{\lambda,i}\sigma^{-1}\map T(\sigma)  \right) \nonumber \\
&=& \sum_{\lambda,i} \lambda c_{\lambda,i}\sigma^{1/2}\left(\sum_k A_k^{\dagger}\sigma^{-1/2}X\sigma^{-1/2}A_k  \right) \sigma^{1/2} \nonumber \\ &=& \sum_{\lambda,i}|\lambda|^2 c_{\lambda,i}X_{\lambda,i}\sigma^{-1/2}\map T^{\dagger}(I) \sigma^{1/2} = \sum_{\lambda,i} c_{\lambda,i} X_{\lambda,i} = X,
\end{eqnarray}
where we use the fact that if $X_{\lambda,i}$ is an attractor associated with eigenvalue $\lambda$ then according to section \ref{sec:monotone_polynom} $\sigma^{-1/2}X_{\lambda,i}\sigma^{1/2}$ is also attractor associated with the same $\lambda$ and satisfying theorem \ref{theorem:attractors of QMCH}. Similarly, one can readily find out $\map T \map T^{\ddagger}(X)=X$. Hence, trace-preserving quantum operation $\map T^{\ddagger}$ constitute a searched generator of the inverse evolution capable to correct an information inscribed into states from the asymptotic space of a given QMP. Moreover, $\map T^{\ddagger}$ is adjoint map of the original generating quantum operation $\map T$ with respect to the scalar product $\left(X,Y\right)_{1/2} = \left(X,\sigma^{-1/2}Y\sigma^{-1/2}\right)$. Thus we can confirm that the asymptotic evolution is unitary, but in a different sense then we are used to. First, it is an unitary evolution on the attractor subspace of operators from the Hilbert space $\set B(\hil H)$, i.e. there is no underlying unitary evolution on the Hilbert space $\hil H$. Second, it is an unitary evolution with respect to a different scalar product on the space $\set B(\hil H)$.

In the case of trace-preserving QMDSs we derive a master equation governing their asymptotic dynamics. Let $X$ be an operator from the attractor space, i.e. according to theorem \ref{theorem:attractors of QMDS} operator $X\sigma^{-1}$ commutes with all Lindblad operators $L_i$'s. The effect of the lindblad generator (\ref{lindbladian_form}) onto $X$ can be simplified as
\begin{eqnarray}
\map L(X) &=&  i[X,H] + \sum_j L_j X\sigma^{-1} \sigma L_j^{\dagger} - \frac{1}{2} \left\{L_j^{\dagger}L_j,X\sigma^{-1}\sigma \right\} \nonumber \\
&=& i[X,H] + X \sigma^{-1} \left(\sum_j L_j X L_j^{\dagger} - \frac{1}{2} \left\{L_j^{\dagger}L_j,X \right\}\right) \nonumber \\
&=& i[X,H] - i X\sigma^{-1}[\sigma,H] = i \left[ HX-X\sigma^{-1}H\sigma \right], \nonumber
\end{eqnarray}
where we use $\map L(\sigma) = 0$. Hence, the master equation takes the form
\begin{equation}
\frac{d\left(X\sigma^{-1} \right)}{dt} = \frac{dX}{dt} \sigma^{-1} = \map L(X) \sigma^{-1} =  i\left[X\sigma^{-1},H\right].
\end{equation}
Thus instead of states, their multiplication with the operator $\sigma^{-1}$ undergo an unitary evolution driven by Hamiltonian $H$. In the same spirit its inverse evolution is driven by Hamiltonian $-H$.

\section{Examples}
\label{sec:examples}
In this part we show two examples demonstrating different aspects of the presented theory. We have chosen two simple but nontrivial examples. The first refers to the creation of entanglement between two qubits and is motivated by the creation of large scale entanglement in a network of many qubits. A full analysis of such a network goes beyond the scope of our paper and is left for a future publication. The other example is motivated by studies of transport of excitation in quantum systems. As in the first example we aim only at illustrating the power of the theory and a complete analysis will be presented elsewhere.


\subsection{Discrete random unitary process}
Let us assume two qubits and two control NOT operations. $U_{12}$ acts on the first qubit as controlled and on the second qubit as target and $U_{21}$ acts in reverse order
\begin{equation}
U_{12}= \left( \begin{array}{cccc}
1 & 0 & 0 & 0 \\
0 & 1 & 0 & 0 \\
0 & 0 & 0 & 1 \\
0 & 0 & 1 & 0 \\
\end{array}
\right), \qquad
U_{21}= \left( \begin{array}{cccc}
1 & 0 & 0 & 0 \\
0 & 0 & 0 & 1 \\
0 & 0 & 1 & 0 \\
0 & 1 & 0 & 0 \\
\end{array}
\right).
\end{equation}
Suppose these two unitary operations act randomly on both qubits with corresponding probabilities $p_{12}$ and $p_{21}$. The resulting propagator determining one step of evolution is a random unitary map
\begin{equation}
\map T(\rho) = p_{12} U_{12} \rho U_{12}^{\dagger} + p_{21} U_{21} \rho U_{21}.
\end{equation}
As this is an unital Markov evolution, it has the same six-dimensional attractor space in both pictures \cite{Novotny2009}. It contains five-dimensional attractor space associated with eigenvalue one, spanned by the identity operator $I$, $|\phi\>\<\phi|$, $|\psi\>\<\psi|$, $|\phi\>\<\psi|$ and $|\psi\>\<\phi|$ with $|\phi\> = |00\>$ and $|\psi\> = 1/\sqrt{3}\left(|01\>+|10\>+|11\> \right)$. Moreover, there is also one dimensional subspace spanned by operator
\begin{equation}
X_{-1}= \left( \begin{array}{cccc}
0 & 0 & 0 & 0 \\
0 & 0 & -1 & 1 \\
0 & 1 & 0 & -1 \\
0 & -1 & 1 & 0 \\
\end{array}
\right)
\end{equation}
associated with eigenvalue $-1$.

Let us assume faithful $\map T$-state $\sigma = 1/5\left(I + |\phi\>\<\phi| \right)$. This $\map T$-state does not commute with all attractors, e.g.  $\left[\sigma,|\phi\>\<\psi| \right] =1/5 |\phi\>\<\psi|$. Consequently, two characterizations of stationary or asymptotic states provided in section \ref{sec:Asymptotic and stationary states of QMPs} are not equivalent. For example, stationary states $\rho_1=1/\mathcal{N}_1 \exp(\log(\sigma) + |\psi\>\<\phi| + |\phi\>\<\psi|)$ and $\rho_2=1/\mathcal{N}_2 \sqrt{\sigma}\exp(|\psi\>\<\phi| + |\phi\>\<\psi|)\sqrt{\sigma}$, with properly chosen normalizations $\mathcal{N}_{1(2)}$, are different.

In order to illustrate that asymptotic states can be obtain as a limit (\ref{limit_exponential_form}) let us assume state
\begin{eqnarray}
\rho = \frac{1}{2}\left(I-|\phi\>\<\phi|-|\psi\>\<\psi| + \frac{1}{\sqrt{3}}X_{-1}\right). \nonumber
\end{eqnarray}
It is asymptotic state which is not strictly positive. By choosing a $\map T$-state proportional to the identity operator, it is easy to verify that it can be written as a limit of strictly positive asymptotic states
\begin{equation}
\rho= \lim_{s \rightarrow +\infty} \frac{1}{\mathcal{N}(s)} \exp\left[\frac{-s}{\sqrt{2}}\left(I-|\phi\>\<\phi|-|\psi\>\<\psi|\right) + \frac{2is}{\sqrt{6}}X_{-1}\right],
\end{equation}
with normalization constant $\mathcal{N}(s)$.

\subsection{Continues QMDS with jump Lindblad operators}
In this example we assume quantum system associated with a four dimensional Hilbert space with orthonormal base $\{|0\>,|1\>,|2\>,|3\> \}$. Its continues Markov evolution is governed by Lindbladian $\mathcal{L}$ which acts as
\begin{equation}
\mathcal{L}(\rho)=-i[H,\rho]+\sqrt{2}\left(h_+\rho h_+^{\dagger}-\frac{1}{2}\{h_+^{\dagger}h_+,\rho\}\right)+h_-\rho h_-^{\dagger}-\frac{1}{2}\{h_-^{\dagger}h_-,\rho\},
\end{equation}
with
\begin{equation}
\label{def_hamiltonian 1}
H=\varepsilon(\ket{2}\bra{2}+\ket{3}\bra{3}),\quad h_+=\ket{0}\bra{1}+\ket{2}\bra{3}=h_-^{\dagger}.
\end{equation}
As it is shown below, the system is equipped with $\mathcal{T}-$state $\sigma$ and thus our developed theory applies. The attractor space contains subspaces corresponding to eigenvalues $0$ and $\pm i\varepsilon$. The former is two-dimensional, spanned by operators $\{X_1, X_2\}$ and each of the latter is one-dimmensional, spanned by operators $X_{\pm}$. These operators read
\begin{align*}
X_1&=2\ket{0}\bra{0}+\ket{1}\bra{1},&&X_2=2\ket{2}\bra{2}+\ket{3}\bra{3},\\
X_+&=2\ket{0}\bra{2}+\ket{1}\bra{3},&&X_-=2\ket{2}\bra{0}+\ket{3}\bra{1}.
\end{align*}

Consequently, the attractor space in the Heisenberg picture is also four-dimensional. It consists of two-dimensional subspace of integrals of motion, which is spanned by the identity operator $I$ and Hamiltonian $H$ and two one-dimensional subspaces corresponding to eigenvalues $\mp i\varepsilon$. These subspaces are spanned by operators $A_{\mp}$, which read
\begin{align*}
A_-=\ket{0}\bra{2}+\ket{1}\bra{3},&&A_+=\ket{2}\bra{0}+\ket{3}\bra{1}.
\end{align*}
As we need hermitian basis of this whole attractor space, it is convenient to define hermitian operators $A_R=\frac{1}{2}(A_++A_-)$ and $A_I=\frac{1}{2i}(A_+-A_-)$.

Let us investigate the structure of the asypmtotic/stationary states in details. As an example, we assume $\sigma=\frac{1}{6}(X_1+X_2)$ and one-parameter class of non-stationary asymptotic states $\rho(s)=\frac{1}{6}(X_1+X_2+sX_++sX_-)$. Apparently $\rho(0)=\sigma$ and $\rho(s)$ is strictly positive for any $s\in(-1,1)$ and non-strictly positive for $s=\pm 1$. According to (\ref{eq_exp_asymptotic states}) one can thus write for any $s\in(-1,1)$
\begin{equation}
\rho(s)=\mathcal{N}(s)\exp\left[\ln\sigma-\beta(s)H-\gamma_R(s)A_R-\gamma_I(s)A_I\right].
\end{equation}
The normalization parameter $\mathcal{N}(s)$ replaces the identity operator (integral of motion) and its corresponding multiplier. Since $\rho(s)$ is balanced in $X_1$ and $X_2$, we get $\beta(s)=0$. Furthermore, $\rho(s)$ is real and thus $\gamma_I(s)=0$. By a straightforward calculation, we get
$\gamma_R(s)=\ln\frac{1+s}{1-s}$. Cases $s=\pm 1$ are resolved via limit procedure
\begin{equation}
\rho(\pm 1)=\lim\limits_{s\rightarrow\pm1}\mathcal{N}(s)\exp\left[\text{ln}(\sigma)-\gamma_R(s)A_R\right]=\lim\limits_{\gamma_R\rightarrow\pm\infty}\mathcal{N}(\gamma_R)\exp\left[\text{ln}(\sigma)-\gamma_R A_R\right].
\end{equation}

Stationary states form a special class of asymptotic states. In this examined case all stationary states can be expressed as a linear combination of operators $X_1$ and $X_2$. Thus all stationary states commute with each other and consequently all strictly positive stationary states can be written in equivalent forms
\begin{equation}
\rho=\mathcal{N}\exp\left[\text{ln}(\sigma)-\beta_{\sigma}H\right]=\mathcal{N}\sigma^{\frac{1}{2}}\exp\left[-\beta_{\sigma}H\right]\sigma^{\frac{1}{2}},
\end{equation}
with $\sigma$ being an arbitrary faithful $\mathcal{T}-$state. As an example, let us take $\sigma=\frac{1}{6}(X_1+X_2)$. By direct calculation, one can show that the strictly positive stationary states can be represented as
\begin{equation}
\exp[\text{ln}(\sigma)-\beta H]=\frac{1}{3+3e^{-\beta\varepsilon}}\left(X_1+e^{-\beta\varepsilon}X_2\right),\ \beta\in\mathbb{R}.
\end{equation}

Let us parametrize the stationary states as $\rho(s)=\frac{1}{3}((1-s)X_1+sX_2)$. This linear combination is strictly positive for $s\in(0,1)$ and non-strictly positive for $s\in\{0,1\}$. For $s\in(0,1)$, we can write

\begin{equation}
\rho(s)=\mathcal{N}(s)\exp\left[\text{ln}(\sigma)-\beta(s)H\right],
\end{equation}

where $\beta(s)=\text{ln}\frac{s}{1-s}$ by direct calculation. Cases $s\in\{0,1\}$ are again resolved via limit procedure
\begin{align*}
\rho(0)&=\lim_{s\rightarrow 0}\mathcal{N}(s)\exp\left[\text{ln}(\sigma)-\beta(s)H\right]=\lim_{\beta\rightarrow -\infty}\mathcal{N}(\beta)\exp\left[\text{ln}(\sigma)-\beta H\right],\\
\rho(1)&=\lim_{s\rightarrow 1}\mathcal{N}(s)\exp\left[\text{ln}(\sigma)-\beta(s)H\right]=\lim_{\beta\rightarrow +\infty}\mathcal{N}(\beta)\exp\left[\text{ln}(\sigma)-\beta H\right].
\end{align*}

Both representations of stationary and asymptotic states (\ref{limit_exponential_form}) and (\ref{eq_exp_asymptotic states}) are in this case equivalent. However, if we switch the Hamiltonian (\ref{def_hamiltonian 1}) off, i.e. we set $H=0$, this statement is not anymore true. The size of the attractor space remains unchanged, but all attractors now correspond to the eigenvalue $\lambda=0$. All asymptotic states are thus stationary states. By making choice $\sigma=\frac{1}{12}(2X_1+2X_2+X_++X_-)$, we can show as an example that the stationary states $\rho_1=\mathcal{N}_1\exp[\text{ln}(\sigma)-\gamma (A_R+A_I)]$ and $\rho_2=\mathcal{N}_2\sigma^{\frac{1}{2}}\exp[-\gamma (A_R+A_I)]\sigma^{\frac{1}{2}}$ are not equal. For instance, $\gamma=-\text{ln}(3)$ yields $\rho_1=\frac{1}{12}(2X1+2X2-i(X_++X_-))$ while $\rho_2=\frac{1}{12}(2X1+2X2+\alpha(X_++X_-))$, with $\text{Re}(\alpha)>0$.

\section{Conclusion}
\label{sec:conclusion}
QMCHs and QMDSs are realistic classes of open system dynamics describing a wide range of processes of significant importance in physics. After introducing the needed formal frame we derived a number of results describing the features of the asymptotic dynamics of quantum discrete and continues, in general trace-nonincreasing, Markov processes. In this way we extend significantly the previously known theory for discrete time quantum Markov processes
and generalize its application also to continues quantum Markov semigroups \cite{Novotny2012}. We formulate and prove basic fundamental theorems concerning the asymptotic dynamics and point out a number of its interesting properties. In particular, based on operator monotone functions a general set of relations between attractors of QMPs in both pictures are revealed and specified for two important cases. Consequently, it provides a dual basis of attractors in the Schrodinger picture and thus significantly simplifies the task of finding the asymptotic dynamics for any initial state. Furthermore we derived equations determining attractors of QMPs in both pictures. We showed that the asymptotic evolution of QMPs has a unitary character if we redefine the relevant Hilbert-Schmidt scalar product. However, we stress that this unitary evolution may not correspond to an unitary evolution on the original Hilbert space of pure states. Moreover, based on the developed theory, two characterizations of asymptotic states are provided, both strongly resembling the form of Gibbs states known from statistical physics. This feature will be the subject of further studies as it points out to an intimate relation between the statistical character of the dynamics and the thermodynamic features of the asymptotic dynamics. Finally, we provide two elementary examples to demonstrate how our theory works. The examples describe the simplest possible nontrivial cases. The chosen models have natural extensions.  However they are more involved and physically more interesting. Their detailed studies will be presented elsewhere as they go clearly beyond the scope of the present paper.

\section*{Acknowledgement}
J. N., J.M. and I. J. have been supported by the Czech Science foundation (GA\v{C}R) project number 16-09824S, RVO 6840770, and Grant Agency of the Czech Technical University in Prague, grant No. SGS16/241/OHK4/3T/14.

\end{document}